\documentclass[aip,author-year,reprint,onecolumn,showpacs,showkeys,amsmath,amssymb]{revtex4-1}

\usepackage{graphicx}
\usepackage{dcolumn}
\usepackage{bm}
\usepackage{color}
\usepackage{natbib}

\newcommand{\dd}{{\rm d}}

\begin{document}

\baselineskip 12pt

\title{Time dependence in large amplitude oscillatory shear: a rheo-ultrasonic study of fatigue dynamics in a colloidal gel}

\author{Christophe~Perge}
\affiliation{Laboratoire de Physique, Universit\'e de Lyon - \'Ecole Normale Sup\'erieure de Lyon - CNRS UMR 5672, 46 All\'ee d'Italie, 69364 Lyon cedex 07, France}

\author{Nicolas~Taberlet}
\affiliation{Laboratoire de Physique, Universit\'e de Lyon - \'Ecole Normale Sup\'erieure de Lyon - CNRS UMR 5672, 46 All\'ee d'Italie, 69364 Lyon cedex 07, France}
\affiliation{UFR de Physique, Universit\'e Claude Bernard Lyon 1, Universit\'e de Lyon, 43 boulevard du 11 novembre, 69100 Villeurbanne, France.}

\author{Thomas~Gibaud}
\affiliation{Laboratoire de Physique, Universit\'e de Lyon - \'Ecole Normale Sup\'erieure de Lyon - CNRS UMR 5672, 46 All\'ee d'Italie, 69364 Lyon cedex 07, France}

\author{S\'ebastien~Manneville}
\affiliation{Laboratoire de Physique, Universit\'e de Lyon - \'Ecole Normale Sup\'erieure de Lyon - CNRS UMR 5672, 46 All\'ee d'Italie, 69364 Lyon cedex 07, France}

\date{\today}

\begin{abstract}
We report on the response of a yield stress material, namely a colloidal gel made of attractive carbon black particles, submitted to large amplitude oscillatory shear stress (LAOStress). At a constant stress amplitude well below its apparent yield stress, the gel displays fatigue and progressively turns from an elastic solid to a viscous fluid. The time-resolved analysis of the strain response, of the Fourier spectrum, and of Lissajous plots allows one to define two different timescales $\tau_w<\tau_f$ associated with the yielding and fluidization of the gel. Coupling rheology to ultrasonic imaging further leads to a local picture of the LAOStress response in which the gel first fails at the walls at $\tau_w$ and then undergoes a slow heterogeneous fluidization involving solid--fluid coexistence until the whole sample is fluid at $\tau_f$. Spatial heterogeneities are observed in both the gradient and vorticity directions and suggest a fragmentation of the initially solidlike gel into macroscopic domains eroded by the surrounding fluidized suspension.
\end{abstract}


\maketitle

\section{Introduction}

Among the huge variety of soft materials, dispersions of colloidal particles are involved in a broad range of applications such as paints, food stuff, printing, etc. [\cite{Russel:1987,Mezzenga:2005,Gibaud:2012a}]. On a more fundamental side, colloidal systems also appear as model systems to study both glass and sol--gel transitions [\cite{Sciortino:2005,Cardinaux:2007,Lu:2008}]. In particular, concentrated colloidal dispersions can be categorized as repulsive or attractive glasses depending on the interaction potential between the particles [\cite{Pham:2002,Tanaka:2004}]. For sufficiently attractive particles, low volume fractions may be enough for clusters to form and percolate through the sample, leading to a ``gel'' structure [\cite{Lu:2008}]. Due to the jammed microstructure in concentrated colloids and to the space-spanning microstructure in colloidal gels, these systems all display elastic properties at rest: the storage modulus $G'$ is always much larger than the loss modulus $G''$ when measured close to equilibrium, e.g. through oscillatory shear of small amplitude (typically a strain amplitude of less than 1\% or a stress amplitude of less than 1~Pa) [\cite{Barnes:1989,Larson:1999}]. 

Besides the equilibrium phase behavior of colloidal systems and their (linear) mechanical properties at rest, their (nonlinear) response to large deformations is also of outstanding interest [\cite{Buscall:1993,Larson:1999,Pham:2006,Pham:2008,Brader:2010,Koumakis:2013}]. Indeed, under large deformations, colloidal particles may rearrange by changing neighbors, clusters may break into smaller aggregates, so that the system is able to flow like a liquid, i.e. $G''$ becomes much larger than $G'$. Although it is virtually involved in any practical application of colloidal systems, this yielding transition (or ``unjamming'' transition) still raises many theoretical and experimental challenges [\cite{Moorcroft:2013}]. At the heart of these difficulties is the issue of the {\it time dependence} of the microstructure. Obviously, the time needed to ``unjam'' and fluidize a colloidal system lying in a kinetically arrested state (and more generally any yield stress material) can be anticipated to become very large depending of the strain amplitude and on the protocol details. On the other hand, when strain or stress is decreased on the fluidized colloidal dispersion, it always takes some time for the microstructure to reform. This leads to so-called ``thixotropic'' or ``rheopectic'' features [\cite{Mewis:2009,Moller:2009b}] and can be captured by measuring the timescales involved in step-up or step-down stress or strain relaxation experiments [\cite{Dullaert:2005b,Dullaert:2006}] or through hysteretic loops in up-down flow curves [\cite{Chen:1992,Divoux:2013,Ovarlez:2013}]. Even more generally, increasingly long timescales are expected for flow dynamics upon approaching the yield point, which can be thought of as the critical point in a dynamical phase transition [\cite{Bocquet:2009,Divoux:2012}]. Such time dependence has recently been illustrated through the existence of diverging timescales under steady shear close to yielding, not only in colloidal gels [\cite{Gopalakrishnan:2007,Gibaud:2010,Sprakel:2011,Lindstrom:2012,Grenard:2013pp}] but also in standard yield stress fluids such as microgels [\cite{Uhlherr:2005,Divoux:2010,Divoux:2011a,Divoux:2012}] and soft thermo-reversible protein gels [\cite{Brenner:2013}]. 

A widespread way to probe yielding in soft materials is to follow the evolution of $G'$ and $G''$ as the oscillation amplitude of the stress or strain is progressively increased above linear response. This generally allows one to estimate the degree of elastic vs viscous response and to infer yield criteria, e.g. from the crossover $G'=G''$ in yield stress materials [\cite{Petekidis:2002,Pham:2006,Smith:2007,Renou:2010,Datta:2011,Koumakis:2012b,Shao:2013,Dimitriou:2013}]. For instance, this approach has been used to unveil a major difference between attractive and repulsive colloidal glasses: while the elastic modulus decreases monotonically in dense hard sphere-like systems, a two-step yielding is observed in attractive glasses, which has been related to the destruction of the ``cages'' made by neighboring attractive particles around a given colloid [\cite{Pham:2006,Laurati:2011}]. 

In the last two decades, rheological measurements based on oscillatory shear within the nonlinear regime have made tremendous progress thanks to the development of new tools in Fourier transform (FT) rheology [\cite{Wilhelm:2002b}] and to other refined mathematical tools [\cite{Cho:2005,Klein:2007,Ewoldt:2008}]. FT rheology allows one to go beyond the measurement of the material response at the fundamental oscillation frequency $f_0$ (characterized by the storage and loss moduli $G'$ and $G''$ extended into the nonlinear regime and usually noted $G'_1$ and $G''_1$) by analyzing higher harmonics in the large amplitude oscillatory stress or strain signals [\cite{Wilhelm:2002b}]. Yet the harmonic content lacks a clear physical interpretation. To overcome this drawback of FT rheology, various approaches based on different decompositions of the nonlinear signal have been proposed which provide valuable physical information on the material behavior during one oscillation cycle [\cite{Cho:2005,Klein:2007,Ewoldt:2008}]. Among them, the orthogonal decomposition in Chebyshev polynomials proposed by \cite{Ewoldt:2008} has the advantages of being unique and directly connected to Fourier coefficients, and of providing measures with a simple physical interpretation in terms of strain hardening, strain softening, etc. For more details, a recent enlightening review on such a {\it large amplitude oscillatory shear} (LAOS) rheology can be found in \cite{Hyun:2011}.

In view of the importance of LAOS rheology both for practical applications and for fundamental problems, it is not surprising that this technique, also recently coined ``LAOStrain'' or ``LAOStress'' in order to distinguish between strain-controlled or stress-controlled oscillations [\cite{Lauger:2010,Dimitriou:2013}], has generated an ever-increasing amount of literature since the early 1990s. Much more unexpected is the fact that only a handful of papers are concerned with the possible effect of time dependence on LAOS measurements. For instance, the academic search engine Web of Knowledge returns about 290 and 200 results since 1990 for searches based on the keywords ``large amplitude oscillatory shear'' and ``Fourier-transform rheology'' respectively, but less than five results are found when either of these keywords is associated with ``time dependence'' or ``thixotropy.'' With the notable exceptions of \cite{Li:2009} in the context of polymers, of \cite{Rogers:2011b} in the context of aging glassy star polymers, and of \cite{Dimitriou:2013pp} in the context of waxy crude oils, it appears that a vast majority of works focusing on LAOS experiments take for granted that oscillations are performed in a ``steady-state'', in the sense that the material properties do not significantly change from one oscillation cycle to another and somehow adapt ``instantly'' to any change in the oscillations. At the very least, this requires that the timescales characteristic of microstructural changes remain much shorter than $1/f_0$ and than any other timescale involved in the LAOS protocol, such as the inverse of the sweep rate when the oscillation amplitude is varied through a continuous ramp or the time interval between steps of constant amplitude, which may indeed become the fastest timescale at low oscillation frequencies. Unfortunately, as recalled above, yielding often goes along with slow dynamics. If the material microstructure evolves on timescales longer than those of the LAOS protocol, these dynamics may somewhat complicate the analysis of the LAOS results and even question the validity of their interpretation.

The purpose of the present paper is twofold. It first aims at attracting the attention of the rheology community on the potential influence of time-dependent microstructure on LAOS experiments performed in attractive colloidal systems and more generally in yield stress materials. Second, we show that crucial complementary information on yielding under oscillatory shear can be gained by combining LAOS with ultrasonic imaging. We mainly focus on time-resolved LAOS experiments under a fixed oscillation amplitude in a colloidal gel known to present ``delayed yielding'', namely a carbon black gel. This corresponds to the limiting case where the material evolves very slowly compared to $1/f_0$ during most of the experiment. The paper is organized as follows. Section~\ref{s.meth} describes our sample and experimental setup. We also briefly recall the LAOStress protocol and analysis as well as the potential problems inherent to LAOStress. Our results are gathered in Sect.~\ref{s.results}. We first use the LAOStress measures as described in \cite{Lauger:2010} and \cite{Dimitriou:2013} to characterize the temporal evolution of the oscillatory response of our colloidal gel. We then report on local ultrasonic measurements which allow us to distinguish between two successive steps in the yielding process of our attractive colloidal system and to interpret these steps in terms of yielding at the walls and heterogeneous bulk fluidization. Finally, Sect.~\ref{s.discuss} provides a short discussion of these results as well as a few perspectives for future work.

\section{Material and methods}
\label{s.meth}

\subsection{Carbon black gels}
Carbon black (CB) particles are colloidal carbonated particles with typical sizes ranging from 80 to 500~nm that result from the partial combustion of hydrocarbon oils [\cite{Waarden:1950,Samson:1987}]. Those particles are widely used in the industry, for instance to reinforce mechanical properties or insure the conductivity of plastic and rubber materials [\cite{Donnet:1993}]. When dispersed in oil, these carbon black particles are weakly attractive, with interactions of typical strength $U\sim 30k_BT$, and aggregate into a space-spanning fractal network [\cite{Trappe:2000,Trappe:2007}]. These colloidal dispersions therefore form gels at low concentrations as described above in the introduction.

Here, we focus on carbon black particles (Cabot Vulcan XC72R, density 1.8) dispersed at 6\%~wt. in a mineral oil (density 0.838, viscosity 20~mPa.s, Sigma Aldrich). In order to use ultrasonic imaging in this optically opaque material, the gel is seeded with 1\%~wt. hollow glass microspheres (Potters Sphericel, mean diameter 6~$\mu$m, mean density 1.1~g.cm$^{-3}$) as already described in \cite{Gibaud:2010} and in \cite{Grenard:2013pp}. Hollow glass microspheres act as ultrasonic contrast agents in an otherwise acoustically transparent dispersion, without altering significantly its rheological properties. This system also has the great practical advantage of being chemically stable and insensitive to evaporation: the same gel can be studied for several weeks provided an adequate pre-shear protocol is used to resuspend the seeding microspheres and rejuvenate the gel prior to each experiment.

When a constant stress is applied from rest in a standard creep experiment, CB gels were shown to display ``delayed yielding'': these gels [\cite{Gibaud:2010,Grenard:2013pp}] as well as other colloidal gels [\cite{Gopalakrishnan:2007,Sprakel:2011,Lindstrom:2012}] first seem to remain solid before undergoing a rather sudden fluidization after a well-defined time $\tau_f$. This strongly time-dependent behavior was modeled in the framework of activated dynamics, predicting quasi-exponential decay of the fluidization time $\tau_f$ with the applied stress $\sigma$ [\cite{Gopalakrishnan:2007,Sprakel:2011,Lindstrom:2012}].

Besides time dependence, \cite{Gibaud:2010} have inferred from one-dimensional ultrasonic imaging that the yielding process of CB gels involves wall slip and spatially heterogeneous fluidization. Very recent investigations have also shown that this delayed yielding is affected by boundary conditions. Indeed, using rough boundaries instead of smooth walls introduces an additional timescale associated with an initial power-law creep regime that ends with the gel failure at the moving wall and whose duration diverges as a power-law as the applied shear stress approaches the yield stress [\cite{Grenard:2013pp}]. In the present paper, we shall use LAOS data analysis together with a newly developed two-dimensional ultrasonic imaging technique in order to complement the LAOStress experiments already briefly reported in \cite{Gibaud:2010}. 

\subsection{LAOStress protocol}
Our rheological measurements are performed in a smooth Couette cell of height 50 mm with a rotating inner bob of radius 48~mm made out of PEEK, and a fixed outer cup of radius 50 mm made out of PMMA. The bob is a hollow cylinder in order to reduce its inertia. This Couette cell is mounted on a stress-imposed rheometer (ARG2, TA Instruments). To ensure an initial reproducible gel state, each measurement is carried out by first pre-shearing the suspension at $500$~s$^{-1}$ then at $-500$~s$^{-1}$ for 20~s each in order to break up any large aggregate. Second, we let the gel rest so that it can reform by applying a zero shear rate for 2~s and a zero shear stress for 10~s. Third, we monitor the linear viscoelastic properties of the gel by measuring the elastic modulus $G'$ and loss modulus $G''$ through oscillatory shear at a low stress amplitude of 0.5~Pa and frequency $f_0=1$~Hz for 60~s. A steady state is reached within a few seconds and we find typical elastic and loss moduli $G'\simeq 1.5\,10^3$~Pa and $G''\simeq 150$~Pa (see Fig.~\ref{fig1}). The sample is then left to rest for 10~s by applying a zero shear stress. Finally, we start the large amplitude oscillatory shear stress experiment at time $t=0$ by imposing oscillations of constant frequency $f_0$ over time. The temperature is controlled by a water circulation around the Couette cell and fixed to $25\pm0.1^\circ$C for all experiments. All the data presented here were obtained with the same oscillation frequency $f_0=1$~Hz. The amplitude of the stress-controlled oscillations is noted $\sigma_0$. In Sect.~\ref{s.ramp} below, we shall only briefly report on an experiment where $\sigma_0$ is swept from small to large values in order to highlight how time dependence arises in a standard ramp protocol. All other experiments will be conducted over time at a fixed $\sigma_0$ and we shall mainly focus on a single specific experiment at $\sigma_0=11$~Pa.

\subsection{Potential problems with LAOStress}
\label{s.pb}
It is important to keep in mind that in a time-dependent material, a number of experimental problems may arise when performing LAOStress experiments even with a single-head stress-controlled rheometer [\cite{Lauger:2010}]. Indeed, the rheometer uses a feedback loop to account for the instrument and bob inertia in order to ensure that the stress applied on the sample is sinusoidal with the prescribed amplitude. When the material structure evolves significantly over a small number of successive oscillations, as can be the case for stress-induced fluidization, the feedback may not be efficient enough for the amplitude to match the commanded value and for stress oscillations to remain sinusoidal, especially for large amplitudes and/or low frequencies and if the feedback involves several full oscillation cycles. The most common problem is that when the sample fluidizes, the stress due to inertia (referred to as ``inertia'' stress) increases sharply while the rheometer keeps the same total stress (also referred to as ``raw'' stress), so that the actual stress felt by the sample (referred to as ``sample'' stress) falls below the commanded value. However, most recent rheometers (including the one used in the present study) use a ``direct strain oscillation'' (DSO) procedure to correct for inertia in real-time within each oscillation cycle, which usually minimizes deviations from the required sinusoidal stress [\cite{Lauger:2010}]. Another issue arises from the fact that our specific rheometer uses a small additional position control superimposed to the stress control to avoid any drift in the mean zero position. This corresponds to a ``soft-lock'' torque which reduces the achieved torque. This torque reduction is usually negligible for high sample stiffnesses but may significantly affect the stress signal as the stiffness of the sample goes down due to progressive fluidization. 

More generally, even for non-thixotropic, time-independent materials in a true steady state, inertia may seriously complicate the analysis of LAOStress data [\cite{Dimitriou:2013}]. In particular, in its standard configuration, our rheometer software provides access only to the total stress waveform. In other words, in the analysis of the time-resolved strain and stress data presented below, the stress signal $\sigma(t)=\sigma_{\rm total}(t)$ contains both the sample stress $\sigma_{\rm sample}(t)$ and the contribution from the geometry and instrument inertia $\sigma_{\rm inertia}(t)$. Assuming that the strain response $\gamma(t)$ is sinusoidal with amplitude $\gamma_0$, the amplitude of the oscillatory inertia stress for our small-gap Couette geometry is:
\begin{equation}
\sigma_{\rm inertia,0}=\frac{e}{2\pi H R_1^3}\,\omega_0^2 I \gamma_0\,,
\end{equation}
where $e=2$~mm is the gap width, $R_1=48$~mm is the bob radius, $H=50$~mm is the height of the cell, $\omega_0=2\pi f_0$, and $I$ is the total effective moment of inertia including that of the instrument and of the geometry [see also the discussion in Appendix~B of \cite{Dimitriou:2013} for the case of a cone-plate geometry]. Therefore, for a commanded stress amplitude $\sigma_0$, the contribution of inertia can be neglected as long as:
$\gamma_0\ll \frac{2\pi H R_1^3}{e \omega_0^2 I}\sigma_0$. In the case of the present experiments, the total inertia is measured to be $I=301~\mu$Nms$^{-2}$, where the main contribution comes from our large custom-made bob ($I_{\rm bob}=282~\mu$Nms$^{-2}$). For $f_0=1$~Hz and $\sigma_0=11$~Pa, the above condition reads $\gamma_0\ll \gamma_0^\star\simeq16$. Although this value should only be taken as a rough estimate (since the strain is no longer sinusoidal much below $\gamma_0^\star$), we can confidently neglect inertia and identify the total stress with the sample stress as long as $\gamma_0\lesssim 1$. As will be shown below in Sect.~\ref{s.cst}, this corresponds to the main part of the experiment, where fluidization remains spatially limited. We have also checked that the stress signal remains sinusoidal and that its amplitude matches the commanded stress to within better than 5\% over the same period of time (see Sect.~\ref{s.ft} below).

Finally, knowing the strain waveform $\gamma(t)$, the sample stress can in principle be recovered from the total stress by differentiating $\gamma(t)$ twice and by using:
\begin{equation}
\sigma_{\rm sample}(t)=\sigma_{\rm total}(t)-\sigma_{\rm inertia}(t)=\sigma_{\rm total}(t)-\frac{e}{2\pi H R_1^3}\, I \ddot\gamma(t)\,.
\label{e.sigmacorr}
\end{equation}
However, in the present work, we chose to stick with the rheometer measurement $\sigma_{\rm total}(t)$ for the following reasons: (i)~taking the second derivative of $\gamma(t)$ strongly enhances experimental noise, (ii)~LAOStress measures are computed from $\gamma(t)$ (see Sect.~\ref{s.analysis} below) and are insensitive to the correction in Eq.~(\ref{e.sigmacorr}), and (iii)~even more importantly, we shall show through local measurements that the strain field becomes spatially heterogeneous as fluidization proceeds and the contribution of inertia accordingly increases. In the event of spatial heterogeneity, the LAOStress framework gets clearly questionable and the corresponding measures become hardly interpretable.

\subsection{LAOStress data analysis}
\label{s.analysis}

\subsubsection{Fourier transform rheology}
Besides the usual measurements at the fundamental oscillation frequency that are directly provided by the rheometer ($G'=G'_1$, $G''=G''_1$, and the phase shift $\delta=\delta_1$ between the input stress and the strain response at $\omega_0$), we use standard tools from Fourier transform rheology to compute the full complex Fourier spectrum $J^\star(\omega)$ of the strain response $\gamma(t)$ and the discrete Fourier components $J^\star_n=J^\star(n\omega_0)$, whose real and imaginary parts will be noted respectively $J'_n$ and $J''_n$, with $n\in \mathbb{N}$.

As we shall check below in Fig.~\ref{fig3}, odd harmonics strongly dominate in the strain response. Therefore, following \cite{Dimitriou:2013} and using the adequate normalization for the Fourier coefficients, $\gamma(t)$ can be written as:
\begin{equation}
\gamma(t)=\sigma_0\sum_{n~{\rm odd}}\left\{ J'_n\cos(n\omega_0 t) +  J''_n\sin(n\omega_0 t)\right\}\,,
\label{e.sumodd}
\end{equation}
where $\sigma(t)=\sigma_0\cos(\omega_0 t)$ is the input commanded stress.

As a crude way to quantify nonlinearity, we define the total harmonic distortion (THD) of the strain signal as: 
\begin{equation}
{\rm THD}(\gamma)=\frac{\sqrt{\displaystyle \sum_{n~{\rm odd}~\ge 3}\left\vert J^\star_n\right\vert^2}}{\left\vert J^\star_1\right\vert}\,.
\label{e.thd}
\end{equation}
In order to test for the presence of harmonics in the input stress signal, we will similarly define a total harmonic distortion THD($\sigma$) based on the Fourier coefficients of $\sigma(t)$ (see Fig.~\ref{fig4}).

Since we are interested in time dependence, the Fourier transform is performed in a time-resolved fashion by analyzing successive waveforms recorded over one period by the rheometer during long experiments where the oscillation amplitude and frequency are kept constant. As expected in a time-dependent material, the Fourier coefficients will themselves depend slowly on time, so that they should in principle be noted $J'_n(t_0)$ and $J''_n(t_0)$, where $t_0$ is the slow timescale at which the Fourier transform is performed. In the following however, we shall use the same symbol $t$ for denoting the time within one oscillation cycle and over the whole experiment. The same comment holds for the THD.

\subsubsection{LAOStress measures}
\label{s.laosmeas}
LAOStress data are further analyzed as described in \cite{Lauger:2010} and \cite{Dimitriou:2013}. Here, we only briefly recall the definitions of the various LAOStress measures based on the rheological measurements and their interpretations. First, from the compliance coefficients $J'_n$, two different {\it elastic} measures are introduced: (i)~the zero-stress elastic compliance $J'_M$, defined as:
\begin{equation}
J'_M\equiv \left. \frac{\dd\gamma}{\dd\sigma}\right |_{\sigma=0}=\sum_{n~{\rm odd}}(-1)^{(n-1)/2}\,nJ'_n=J'_1-3J'_3+5J'_5-7J'_7+\cdots\,,
\label{e.JM}
\end{equation}
and (ii)~the large-stress elastic compliance $J'_L$, defined as:
\begin{equation}
J'_L\equiv \left. \frac{\gamma}{\sigma}\right | _{\sigma=\sigma_0}=\sum_{n~{\rm odd}}J'_n=J'_1+J'_3+J'_5+J'_7+\cdots\,.
\label{e.JL}
\end{equation}
These two compliances can be geometrically interpreted by considering the classical Lissajous plot of the strain response $\gamma(t)$ vs the input shear stress $\sigma(t)$ : $J'_M$ and $J'_L$ respectively correspond to the slope of the tangent to the Lissajous plot at $\sigma=0$ and to the slope of the line joining both extremities of the cycle at $\sigma=\pm\sigma_0$. This is illustrated in Fig.~\ref{fig5}(a,c,e).

Similarly, one introduces two {\it viscous} measures based on the shear rate response $\dot\gamma(t)$: (i)~the zero-stress fluidity $\phi'_M$, defined as:
\begin{equation}
\phi'_M\equiv \left. \frac{\dd\dot\gamma}{\dd\sigma}\right |_{\sigma=0}=\sum_{n~{\rm odd}}(-1)^{(n-1)/2}\,n^2\omega_0J''_n=\omega_0(J''_1-9J'_3+25J''_5-49J''_7+\cdots)\,,
\label{e.phiM}
\end{equation}
and (ii)~the large-stress fluidity $\phi'_L$, defined as:
\begin{equation}
\phi'_L\equiv \left. \frac{\dot\gamma}{\sigma}\right | _{\sigma=\sigma_0}=\sum_{n~{\rm odd}}n\omega_0J''_n=\omega_0(J''_1+3J''_3+5J''_5+7J''_7+\cdots)\,.
\label{e.phiL}
\end{equation}

Equations~(\ref{e.phiM}) and (\ref{e.phiL}) simply correspond to Eqs~(\ref{e.JM}) and (\ref{e.JL}) where the shear rate $\dot\gamma$ is used instead of the strain $\gamma$ and the fluidity coefficients $n\omega_0 J''_n$ are used instead of the compliance coefficients $J'_n$. Consequently, $\phi'_M$ and $\phi'_L$ have a similar geometric interpretation but for Lissajous plots of $\dot\gamma(t)$ vs $\sigma(t)$ as shown in Fig.~\ref{fig5}(b,d,f).

$J'_M$ and $\phi'_M$ characterize the nonlinear behavior at the minimum stress $\sigma=0$, while $J'_L$ and $\phi'_L$ quantify nonlinearity at the maximum stress $\sigma=\pm\sigma_0$. In order to grasp the intra-cycle elastic behavior with a single measure, \cite{Dimitriou:2013} proposed to consider the relative ratio $R$ of the change in compliance within a LAOStress cycle:
\begin{equation}
R\equiv \frac{J'_L-J'_M}{J'_L}=\frac{4J'_3 - 4J'_5 + 8J'_7 - 8J'_9+\dots}{J'_1+J'_3+J'_5+\cdots}\,.
\label{e.R}
\end{equation}
A large positive value of $R$ corresponds to pronounced intra-cycle stress softening while a negative value indicates stress stiffening. Therefore, $R$ is referred to as the ``stress-softening index.'' Finally, a similar ratio $Q$, referred to as the ``stress-thinning index,'' is defined for the intra-cycle change of fluidity:
\begin{equation}
Q\equiv \frac{\phi'_L-\phi'_M}{\phi'_L}\simeq \frac{12J''_3 - 20J''_5 + 56J''_7 -72J''_9 +\dots}{J''_1+3J''_3+5J''_5+\cdots}\,,
\label{e.Q}
\end{equation}
with $Q>0$ (resp. $Q<0$) indicating stress-thinning (resp. stress-thickening) behavior within an oscillation cycle. As already noted above for the Fourier coefficients and for the THD, all the various LAOStress measures introduced in this section are expected to slowly vary with time in our time-dependent colloidal gel.

\subsection{Ultrasonic imaging}
\label{s.usvsetup}

\begin{figure}[h]
\centerline{\includegraphics[width=0.6\columnwidth]{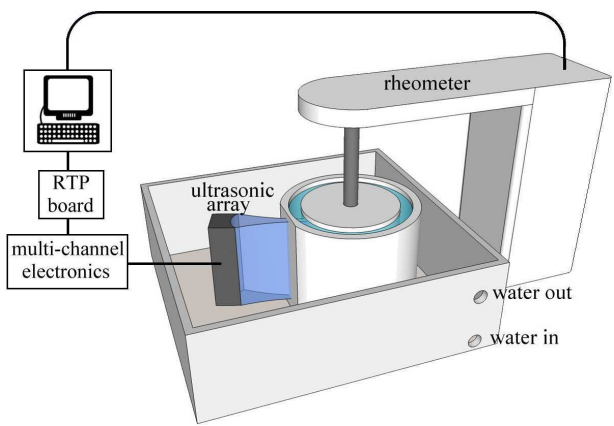}}
\caption{Sketch of the experimental arrangement showing the Couette cell coupled to the rheometer and the ultrasonic transducer array used for imaging the local displacement of the sample from one oscillation cycle to another.}
\label{fig0}
\end{figure}

Just like standard rheology provides information that can be reliably interpreted provided the sample and the strain/flow field remain homogeneous, the LAOS measures introduced above are meant to describe homogeneous situations only. For instance, it has been shown that wall slip may lead to the presence of even harmonics in LAOS signals and in some cases to nonperiodic responses [\cite{Graham:1995}]. Of course, the interpretation of LAOS measures becomes even more problematic in the case of bulk solid--fluid coexistence such as that frequently encountered in yield stress materials under steady shear. In order to test for strain heterogeneity in our CB gel, we use ultrasonic imaging as already introduced by \cite{Gibaud:2010}. Here we adapt the previous one-dimensional technique to two-dimensional imaging based on a newly developed multi-channel setup.

Our ultrasonic setup is sketched in Fig.~\ref{fig0}. It has been described at length in \cite{Gallot:2013} to which the reader is referred for full details. In brief, our technique relies on the use of an array made of 128 piezoelectric transducers that emit and receive low-intensity ultrasonic pulsed waves with a central frequency of 15~MHz. The total length of the transducer array is 32~mm. The array is set vertically (i.e. along the vorticity direction $z$) in a large water tank surrounding the Couette cell at about 30~mm from the outer wall of the cup. A plane pulse is emitted by firing all 128 transducer elements simultaneously. This pulse gets scattered by the hollow glass microspheres seeding the CB gel and the backscattered signals are received by each of the 128 transducers and stored on a personal computer through a real-time processing (RTP) board. From the 128 backscattered signals, an ultrasonic image is formed and yields a ``speckle image'' $S(r,z,t)$ of the scatterer distribution across the gap of the Couette cell (with $r$ the radial distance to the inner bob and $z$ the vertical position from the top of the transducer array) at the time $t$ when the pulse was sent (neglecting the ultrasonic travel time from/to the transducer array which amounts to about 40~$\mu$s.).

The spatial resolution in the radial direction is roughly given by the ultrasonic wavelength 100~$\mu$m and the spacing of 250~$\mu$m between two adjacent transducers on the array gives the vertical resolution. A much finer resolution on the scatterer displacements (down to a few microns) is obtained by cross-correlating successive images $S(r,z,t)$ and $S(r,z,t+\delta t)$ as described in \cite{Manneville:2005} and \cite{Gallot:2013}. Due to the limited spatial resolution along $z$ and since the displacements are expected to be mainly horizontal, the correlation is performed only on the horizontal direction $r$. Under a constant shear stress (or strain rate), this leads to images of the displacement field $\Delta(r,z,t)$ that are directly proportional to the velocity field $v(r,z,t)$ projected along the acoustic propagation axis. In Sect.~\ref{s.usv} below, we set the time interval $\delta t$ to be a multiple of the oscillation period in order to focus on the displacement of the sample between different cycles and emphasize the slow time dependence of our colloidal gel under LAOS. 

\section{Results}
\label{s.results}

\subsection{Ramp experiment}
\label{s.ramp}

\begin{figure}[h]
\centerline{\includegraphics[width=0.6\columnwidth]{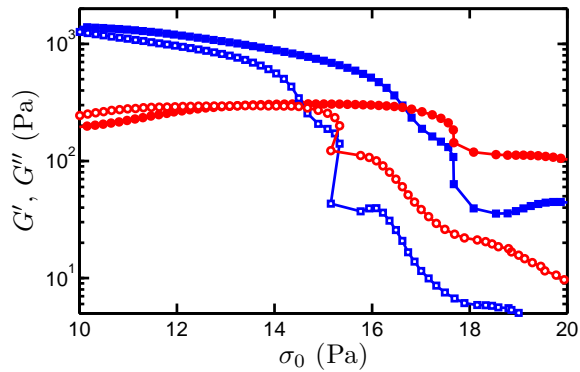}}
\caption{Standard LAOStress sweep experiment illustrating the yielding and fluidization of a $6\%$~wt. carbon black gel at frequency $f_0=1$~Hz. Elastic modulus $G'$ (blue) and loss modulus $G''$ (red) measured when sweeping up the oscillatory stress amplitude $\sigma_0$ for two different sweep rates: 34~mPa.s$^{-1}$ (filled symbols) and 7~mPa.s$^{-1}$ (open symbols).}
\label{fig1}
\end{figure}

Our first experimental results are reported in Fig.~\ref{fig1} and concern a standard oscillation procedure where the imposed stress amplitude is linearly swept up starting from the linear regime up to a fully nonlinear state at a fixed frequency $f_0=1$~Hz. Shown with filled symbols are the viscoelastic moduli $G'$ (in blue) and $G''$ (in red) recorded by the rheometer as a function of the stress amplitude $\sigma_0$ for a ``fast'' sweep rate of 34~mPa.s$^{-1}$. Initially, the gel displays solidlike behavior with $G'\gg G''$. It then apparently yields ($G'\simeq G''$) when $\sigma_0$ reaches a characteristic yield value $\sigma_y\simeq 16.6$~Pa. Both moduli further present a marked shoulder at even larger stress amplitudes before reaching a truly fluidlike state where $G''\gg G'$. 

Surprisingly, if one uses a slower sweep rate of 7~mPa.s$^{-1}$, $G'$ and $G''$ (open symbols) show a similar yet much sharper stress dependence and cross at $\sigma_y\simeq14.6$~Pa. Therefore, the material response to an upward sweep in stress oscillation amplitude strongly depends on the ramp velocity. Such a dependence on the sweep rate is characteristic of thixotropic materials as more commonly demonstrated through hysteretic behaviors in flow curve measurements [\cite{Mewis:2009,Divoux:2013}]. Note that the present result is incompatible with an aging process since both $G'$ and $\sigma_y$ are expected to increase in a material subject to structural consolidation [\cite{Negi:2010b}]. Here, the same CB gel rather yields at a lower stress when submitted to similar stresses for longer times. It can be shown through a more systematic investigation that the apparent yield stress consistently decreases as the sweep rate is decreased. This behavior hints at a ``fatigue'' process under oscillatory shear. Consequently, the evolution of the viscoelastic moduli shown in Fig.~\ref{fig1}, in particular the various steps and drops in $G'$ and $G''$, cannot be interpreted in terms of a steady-state response and these measurements call for a temporally resolved picture of the yielding dynamics under LAOS.

As a final note, we emphasize that the data points in Fig.~\ref{fig1} are not equally spaced with $\sigma_0$ although a linear sweep is imposed. For the slower sweep, the stress amplitude even decreases for $\sigma_0\simeq 15$~Pa. This problem is typical of the difficulties to reach the commanded stress amplitude in a strongly time-dependent material and results from both the increasing inertia and the feedback loop of the rheometer as mentioned above in Sect.~\ref{s.pb}. 

\subsection{Constant amplitude experiment: LAOS stress rheology}
\label{s.cst}

\begin{figure}[h]
\centerline{\includegraphics[width=0.8\columnwidth]{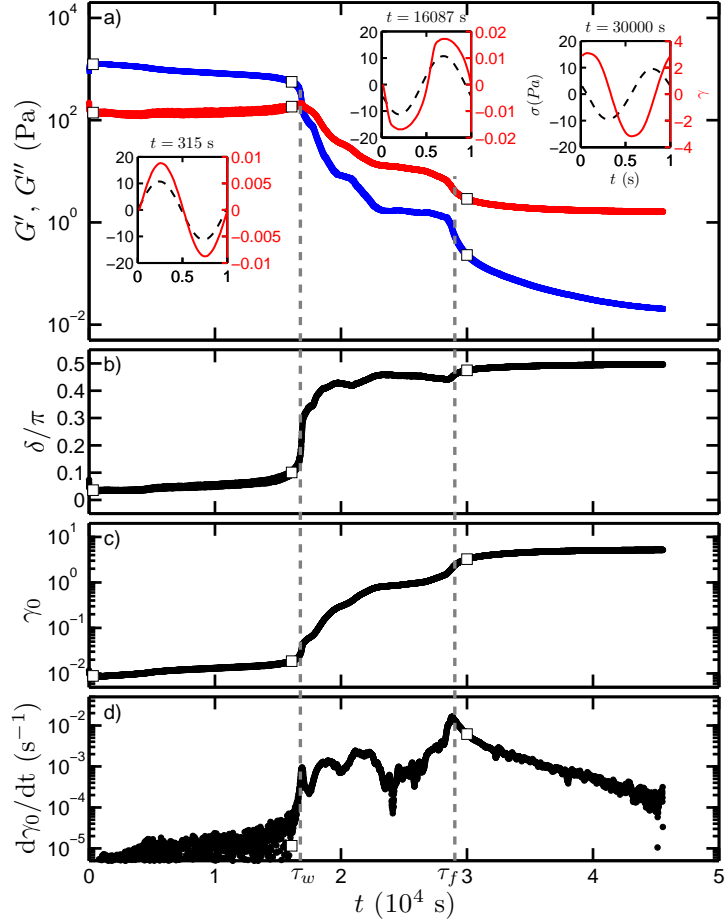}}
\caption{Time-resolved LAOStress experiment illustrating the yielding and fluidization of a $6\%$~wt. carbon black gel under an oscillatory stress of constant amplitude $\sigma=11$~Pa and frequency $f_0=1$~Hz. (a)~Time evolution of the elastic modulus $G'$ (blue) and loss modulus $G''$ (red). Initially, the gel displays solidlike behavior ($G'\gg G''$), apparently yields ($G'\simeq G''$) at $t\simeq1.6\,10^4$~s, and seems to flow like a liquid thereafter ($G'\ll G''$). Insets: ``instantaneous'' strain response $\gamma(t)$ (red solid curves) as a function of the imposed sinusoidal shear stress $\sigma(t)$ (black dashed curves) over one oscillation period at three different times indicated by square symbols in the main graph. (b)~Time evolution of the phase shift $\delta=\tan^{-1}(G''/G')$ between $\sigma(t)$ and $\gamma(t)$. (c)~Time evolution of the strain amplitude $\gamma_0$. (d)~Time derivative ${\rm d}\gamma_0/{\rm d}t(t)$ of the strain amplitude response shown in (c). Gray dashed lines indicate the two characteristic times discussed in the text: the time $\tau_w$ such that $G'(\tau_w)=G''(\tau_w)$, which defines apparent yielding, and the time $\tau_f$ at which ${\rm d}\gamma_0/{\rm d}t(t)$ reaches a global maximum, which corresponds to full fluidization as inferred from ultrasonic imaging.}
\label{fig2}
\end{figure}

For the rest of this paper, we shall mainly focus on a single LAOStress experiment performed at a constant stress amplitude $\sigma_0=11$~Pa and a constant frequency $f_0=1$~Hz over more than $4\,10^4$~s. This experiment is performed at a sufficiently low stress amplitude, so that the various steps of the yielding dynamics can be easily distinguished. We first discuss the standard rheological measurements performed by the rheometer and then turn to Fourier transform rheology and LAOStress measures.

\subsubsection{Standard rheological measurements}
Figure~\ref{fig2}(a) shows the temporal evolution of the viscoelastic moduli $G'$ and $G''$ together with examples of stress and strain waveforms over one oscillation cycle taken at three different stages of the experiment. Figure~\ref{fig2}(b) and \ref{fig2}(c) respectively report the phase shift $\delta$ of the strain response relative to the stress input and the strain amplitude $\gamma_0$ as recorded by the rheometer as a function of time. Here, the phase shift $\delta$ is computed based on the sample stress while the stress waveforms in Fig.~\ref{fig2}(a) are raw signals. These standard rheological data clearly reveal a strong time dependence of the colloidal gel under study. Although $\sigma_0$ lies below the apparent yield stresses inferred from both the amplitude sweeps of Fig.~\ref{fig1}, the CB gel slowly turns from an elastic solid to a viscous liquid if given enough time to yield. Indeed, one has $G'\gg G''$ and $\delta\simeq 0$ for $t\lesssim 1.5\,10^4$~s while $G'\ll G''$ and $\delta\simeq \pi/2$ for $t\gtrsim 3\,10^4$~s (see also the sinusoidal waveforms at $t=315$ and $3\,10^4$~s that are respectively in phase and in phase quadrature). Meanwhile, the strain amplitude $\gamma_0$ increases from about 0.01 by more than two orders of magnitude to reach an asymptotic value of about 5. 

From the slow evolution of the strain amplitude $\gamma_0$ shown in Fig.~\ref{fig2}(c), one may compute the time derivative $\dd\gamma_0/\dd t$, which should not be confused with the amplitude of the oscillatory shear rate $\dot\gamma_0=\omega_0\gamma_0$. The temporal evolution of $\dd\gamma_0/\dd t$ over the whole experiment is plotted in Fig.~\ref{fig2}(d). After a long period of time where it takes very small values, $\dd\gamma_0/\dd t$ goes through a first local maximum, which nicely corresponds to the point at which $G'$ and $G''$ coincide (and therefore $\delta=\pi/4$). This defines a first timescale $\tau_w$ associated with the yielding dynamics of a CB gel under LAOStress and which we will refer to as ``apparent yielding.'' In the present case, one has $\tau_w=16910\pm 10$~s and $\gamma_0\simeq 0.04$ at $\tau_w$. At this point, the strain response has become strongly nonlinear as can be checked from the waveform at $t=16087$~s in the inset of Fig.~\ref{fig2}(a).

Furthermore, a second characteristic timescale $\tau_f$ can be inferred from Fig.~\ref{fig2}(d), which corresponds to the global maximum reached by $\dd\gamma_0/\dd t$, here for $\tau_f=28950\pm 50$~s. Between $\tau_w$ and $\tau_f$, the rheological response looks very complex. From the fact that $G''>G'$, the sample appears fluidlike. The general trend is that both viscoelastic moduli decrease sharply while $\gamma_0$ strongly increases, but the details are rather bumpy and fluctuating as can be seen from $\dd\gamma_0/\dd t$. When $\tau_f$ is reached, $\gamma_0\simeq 2.3$ and all rheological quantities $G'$, $G''$, $\delta$, and $\gamma_0$ eventually enter a much smoother final relaxation towards steady state.

The temporal behavior of our gel characterized by two characteristic timescales is reminiscent of the response of attractive glasses in LAOStrain sweep experiments by \cite{Pham:2006} and \cite{Laurati:2011} who have defined two successive yielding processes as a function of strain amplitude: a first one at $\gamma_{1}=0.01$--0.05 interpreted as the stretching of the bonds between the colloids and a second one at $\gamma_{2}=1$--3 interpreted as the dislocation of the cages formed by the neighboring colloids. Although the order of magnitudes of $\gamma_0(\tau_w)$ and $\gamma_0(\tau_f)$ are strikingly consistent with the two characteristic strains $\gamma_{y1}$ and $\gamma_{y2}$ seen in LAOStrain, it is however precipitate to draw any conclusion in the absence of more local insights. In order to investigate the LAOStress response in more details, we turn to nonlinear analysis in the next two paragraphs.

\subsubsection{Fourier transform rheology}
\label{s.ft}

\begin{figure}[h]
\centerline{\includegraphics[width=0.5\columnwidth]{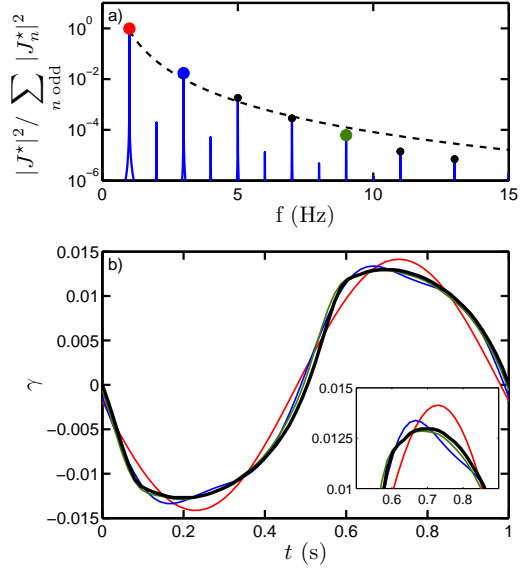}}
\caption{Fourier transform rheology of the LAOStress experiment shown in Fig.~\ref{fig2} at time $t=16087$~s. (a)~``Instantaneous'' normalized power spectrum of the strain response $\gamma(t)$ shown in the middle inset of Fig.~\ref{fig2}(a). The dashed line shows the power-law behavior $\vert J^\star(\omega)\vert\sim 1/\omega^2$. (b)~Reconstruction of the experimental signal (thick black line) at time $t=16087$~s based on the Fourier transform of the instantaneous strain response $\gamma(t)$ with different odd harmonic contents: fundamental only (red), fundamental and third harmonic (blue), and up to the 9$^{\rm th}$ harmonic (green).}
\label{fig3}
\end{figure}

Figure~\ref{fig3}(a) shows the power spectrum $\vert J^\star(\omega)\vert^2$ of the strain response at time $t=16087$~s already displayed in the middle inset of Fig.~\ref{fig2}(a) and replotted in Fig.~\ref{fig3}(b) with a thick black line. This particular time was purposely chosen slightly below $\tau_w$ so as to correspond to a time when the harmonic content is among the largest. The power spectrum is normalized by the contribution of the odd harmonics $\sum_{n~{\rm odd}}\vert J^\star_n\vert^2$. The apparent non-zero width of the fundamental peak and of the third harmonic results from the windowing used for the Fourier transform. We can check in Fig.~\ref{fig3}(a) that odd harmonics dominate in the power spectrum up to the 9$^{\rm th}$ harmonic. Therefore, in the following, summation over odd harmonics will always be restricted to $n=9$. Indeed, even though the second harmonic $\vert J_2^\star\vert$ is two orders of magnitude smaller than the fundamental $\vert J_1^\star\vert$ and one order of magnitude smaller than the third harmonic $\vert J_3^\star\vert$, even harmonics become comparable to odd harmonics with $n\ge 9$. This non-negligible content in even harmonics may be related to a small amount of wall slip around $\tau_w$ as will be further discussed in Sect.~\ref{s.usv} below. Figure~\ref{fig3}(a) also shows that $\vert J_n^\star\vert$ roughly decreases as $1/n^2$ which is typical of a triangular signal. Finally, considering only the odd harmonics up to $n=9$ allows one to reconstruct accurately the strain response as shown by the green line in Fig.~\ref{fig3}(b).

\begin{figure} [h]
\centerline{\includegraphics[width=0.7\columnwidth]{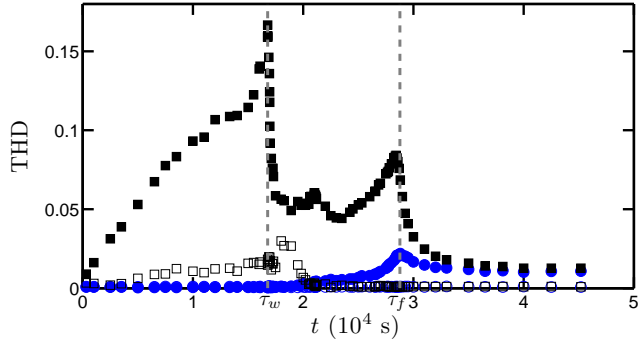}}
\caption{Time evolution of the total harmonic distortion of the strain response THD$(\gamma)$  computed using Eq.~(\ref{e.thd}) over odd harmonics up to $n=9$ ($\blacksquare$) and of the stress signal THD$(\sigma)$ (blue $\bullet$). Open symbols show the THD computed over even harmonics up to $n=8$ for the strain response ($\square$) and for the stress signal (blue $\circ$). Gray dashed lines indicate $\tau_w$ and $\tau_f$. Same experiment as in Fig.~\ref{fig2}.} 
\label{fig4}
\end{figure}

In Fig.~\ref{fig4}, we plot the total harmonic distortion of the strain response THD$(\gamma)$ defined by Eq.~(\ref{e.thd}) as a function of time over the whole experiment corresponding to Fig.~\ref{fig2}. The three different regimes unveiled in the previous section appear even more clearly on THD$(\gamma)$. For $t<\tau_w$, the odd harmonics are seen to grow almost linearly with time up to about 15\% (see black squares): as the amplitude of the strain slowly increases, nonlinearity builds up in the initially elastic material which departs from the linear regime right from the start of the LAOStress oscillations. The contribution of even harmonics to the spectrum remains limited (below 2\%, see empty squares) in this initial phase although it becomes measurable as soon as $t\gtrsim 5\,10^3$~s pointing to possible wall slip. Apparent yielding at $\tau_w$ coincides with a strong peak in THD$(\gamma)$, which falls off very quickly down to about 5\% for $t\gtrsim \tau_w$. As already noted above, the intermediate times $\tau_w<t<\tau_f$ are characterized by complex fluctuations. We also note that the even harmonics become comparable to the odd harmonics for $\tau_w<t\lesssim 2\, 10^4$~s, which hints to important effects of wall slip as will be confirmed below through ultrasonic imaging. Finally, THD$(\gamma)$ goes through a final well-defined maximum that coincides with $\tau_f$. Interpreting this maximum is however tricky since at this stage of the yielding process, the total stress signal includes a significant contribution from inertia. Indeed, the strain amplitude exceeds the rough criterion $\gamma_0\simeq 1$ obtained in Sect.~\ref{s.pb} for $t\gtrsim 2.5\,10^4$~s. This corresponds to the point where THD$(\sigma)$ computed over the odd harmonics (see filled blue circles in Fig.~\ref{fig4}) starts to increase significantly and then passes through a maximum at $\tau_f$. Note that the even harmonics of $\sigma(t)$ (empty circles) remain completely negligible throughout the experiment while those of $\gamma(t)$ (empty squares) only get fully negligible for $t\gtrsim 2\,10^4$~s. We also checked that in this final regime, the amplitude of the stress signal progressively decays from the commanded value, eventually reaching about 9~Pa instead of 11~Pa (data not shown). Together with the fact that asymptotically THD$(\gamma)\simeq\;$THD$(\sigma)$, this clearly shows that the steady state reached for $t>\tau_f$ is a fluidlike state where inertia dominates the rheological recordings.  

\subsubsection{Lissajous plots and LAOStress measures}

\begin{figure} [h]
\centerline{\includegraphics[width=0.8\columnwidth]{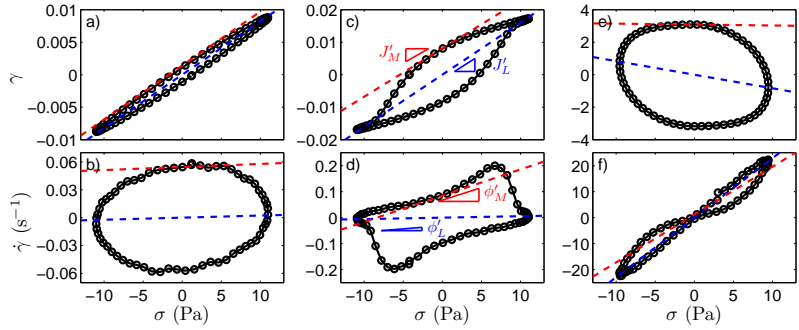}}
\caption{``Instantaneous'' Lissajous plots [$\sigma(t)$, $\gamma(t)$] (top) and [$\sigma(t)$, $\dot\gamma(t)$] (bottom) at (a,b)~$t=315$~s, (c,d)~$t=16087$~s, and (e,f)~$t=3\,10^4$~s corresponding to the waveforms shown in the insets of Fig.~\ref{fig2}(a). Red dashed lines show the slopes $J'_M$ (top) and $\phi'_M$ (bottom). Blue dashed lines indicate $J'_L$ (top) and $\phi'_L$ (bottom). See also Supplementary Movie~1 for the full data set.}
\label{fig5}
\end{figure}

\begin{figure} [h]
\centerline{\includegraphics[width=0.8\columnwidth]{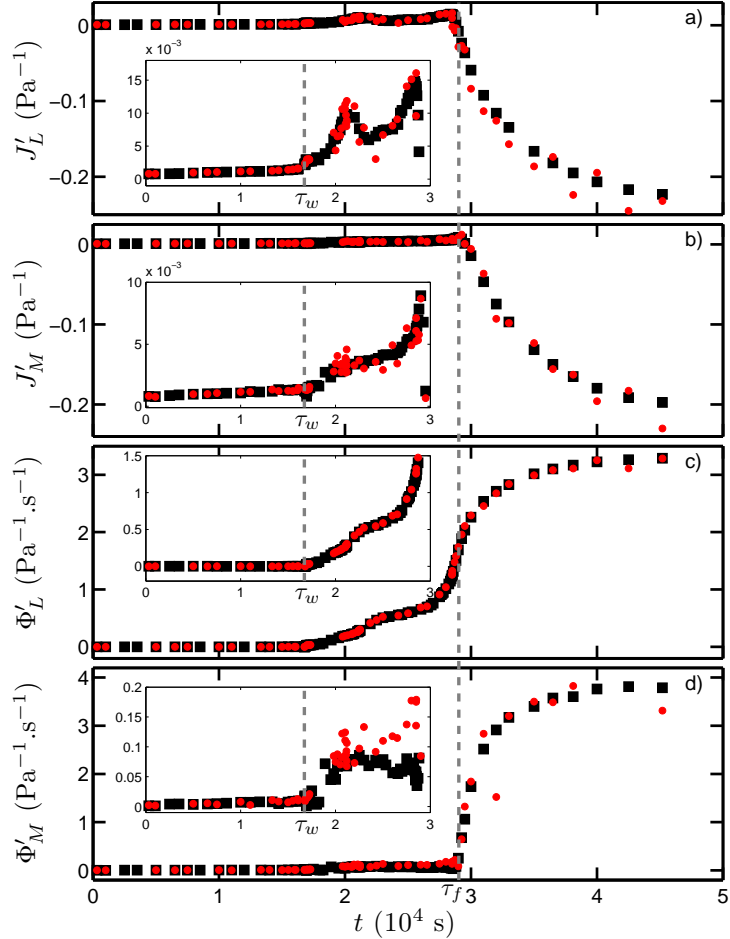}}
\caption{Time evolution of the LAOStress measures defined in Eqs.~(\ref{e.JM})--(\ref{e.phiL}): (a)~$J'_L$, (b)~$J'_M$, (c)~$\phi'_L$, and (d)~$\phi'_M$. Black squares correspond to the estimation based on the Fourier coefficients while red circles are estimated from the slopes in the Lissajous plots. Insets are enlargements for $t<\tau_f$. Gray dashed lines indicate $\tau_w$ and $\tau_f$. Same experiment as in Fig.~\ref{fig2}.}
\label{fig6}
\end{figure}

The temporal build-up of nonlinearity and the subsequent transition to a fluidized state can also be directly visualized through the Lissajous curves of Fig.~\ref{fig5} (see also Supplementary Movie~1). There, the three waveforms shown in Fig.~\ref{fig2}(a) and representative of the various yielding steps are graphed as strain $\gamma$ vs stress $\sigma$ parametrized by time $t$ (top row in Fig.~\ref{fig5}). The bottom row reports the shear rate $\dot\gamma(t)$ [computed from $\gamma(t)$] as a function of $\sigma(t)$. The initial ellipses [Fig.~\ref{fig5}(a,b)] point to quasi-linear viscoelastic response and the fact that the semiminor axis is much smaller than the semimajor axis indicates the predominance of elasticity over viscosity (also note the change in vertical scales from one set of Lissajous plots to the other). Close to apparent yielding at $\tau_w$, the cycle gets strongly distorted due to nonlinearity [Fig.~\ref{fig5}(c,d)]. Eventually, the Lissajous plot recovers an elliptical shape yet now dominated by viscosity [Fig.~\ref{fig5}(e,f)]. Note that the phase shift (i.e. the orientation of the semimajor axis relative to the horizontal) in Fig.~\ref{fig5}(e) becomes larger than $\pi/2$. This is because the stress signal corresponds to the total stress, which gets dominated by the inertial term in Eq.~(\ref{e.sigmacorr}) at the late stages of the experiment. Since the inertial term involves a second derivative of $\gamma(t)$, it corresponds to a phase shift of $\pi$ which explains the ellipse orientation and the negative slope for $J'_L$ in Fig.~\ref{fig5}(e) as well as the self-intersecting cycle in Fig.~\ref{fig5}(f). 

\begin{figure} [h]
\centerline{\includegraphics[width=0.8\columnwidth]{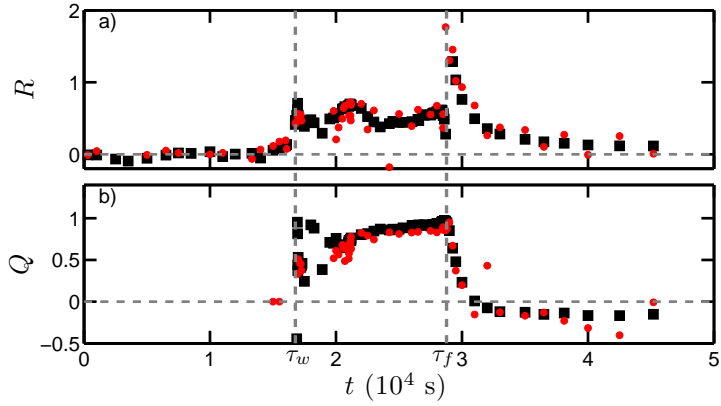}}
\caption{Time evolution of (a) the stress-softening index $R=(J'_L-J'_M)/J'_L$ and (b) the stress-thinnning index $Q=(\phi'_L-\phi'_M)/\phi'_L$  inferred from the data of Fig.~\ref{fig6}. Black squares correspond to the estimation based on the Fourier coefficients while red circles are estimated from the slopes in the Lissajous plots. Gray vertical dashed lines indicate $\tau_w$ and $\tau_f$.}
\label{fig7}
\end{figure}

As recalled in Sect.~\ref{s.laosmeas} and as indicated by dashed lines in Fig.~\ref{fig5}, Lissajous plots can be further analyzed in terms of their slopes at the minimum stress and of the slopes of the lines joining their extremities in order to estimate the various LAOStress elastic ($J'_M$ and $J'_L$) and viscous ($\phi'_M$ and $\phi'_L$) measures. The same measures can be inferred from their expressions in terms of the Fourier coefficients given by Eqs.~(\ref{e.JM})--(\ref{e.phiL}). It can be checked in Figs.~\ref{fig6} and \ref{fig7} that these two different ways of measuring the LAOStress observables as well as the corresponding ratios $R$ and $Q$ are in excellent quantitative agreement. Estimates based on Lissajous plots are naturally subject to more experimental scatter, especially $\phi'_M$ due to the differentiation of the strain signal and to the determination of the tangent at $\sigma=0$. The most prominent feature of the LAOStress measures that appears in Fig.~\ref{fig6} is their time dependence for $t>\tau_f$: both compliances $J'_M$ and $J'_L$ turn negative and then strongly decrease while the fluidities $\phi'_M$ and $\phi'_L$ sharply increase. However, as already evidenced above, the data for $t>\tau_f$ are dominated by the geometry inertia and mix information on the sample behavior with the contribution of the inertia stress. Therefore, although the increase of fluidity is eventually expected for a yielding material, LAOStress measures should only be considered as representative of the sample response for $t<\tau_f$.

The insets in  Fig.~\ref{fig6} shows enlargements of $J'_M$, $J'_L$, $\phi'_M$, and $\phi'_L$ for $t<\tau_f$. It can be seen that the compliances $J'_L$ and $J'_M$ both weakly increase for $t<\tau_w$, which corresponds to the global slow softening of the CB gel already observed in the initial response of the elastic modulus $G'$ [see Fig.~\ref{fig2}(a)]. The corresponding fluidities $\phi'_M$ and $\phi'_L$ remain very close to zero as expected for a solidlike material. Interestingly, the stress-softening index $R$ remains fairly negligible for $t<\tau_w$ [see Fig.~\ref{fig7}(a)]. In other words, the intra-cycle response (or at least the difference in maximum-stress and minimum-stress behaviors) does not show any pronounced stress-thinning nor stress-thickening although nonlinearity progressively builds up for $t<\tau_w$ and reaches a maximum at $t\simeq \tau_w$. Note that for $t<\tau_w$, the stress-thinning index $Q$ cannot be defined since $\phi'_M\simeq\phi'_L\simeq 0$. The time dependence of all LAOStress measures changes markedly at $t\simeq\tau_w$ after which they sharply increase. This reflects the progressive elasticity loss and fluidity build-up in the material. As seen in Fig.~\ref{fig7}, this change of behavior results in values of both indices $R$ and $Q$ of the order of unity for $\tau_w\lesssim t\lesssim \tau_f$. This corresponds to significant apparent stress softening and stress thinning of the CB gel within one oscillation cycle. Here again, for $t>\tau_f$, the predominance of inertia prevents one from drawing any conclusion from the final relaxation of $R$ and $Q$.

\subsection{Constant amplitude experiment: ultrasonic imaging}
\label{s.usv}

The LAOStress analysis presented above provides interesting insights into the delayed yielding of our CB gel. In particular, it allows one to clearly define two different timescales $\tau_w$ and $\tau_f$ characteristic of the yielding dynamics. However, it assumes that the sample remains homogeneous over time. In other words, the above data can only be considered as the result of a spatial average over the whole sample, including boundary layers that may have a crucial influence in the occurrence of apparent wall slip. We now turn to ultrasonic imaging in order to check for the sample homogeneity.

\begin{figure} [h]
\centerline{\includegraphics[width=1\columnwidth]{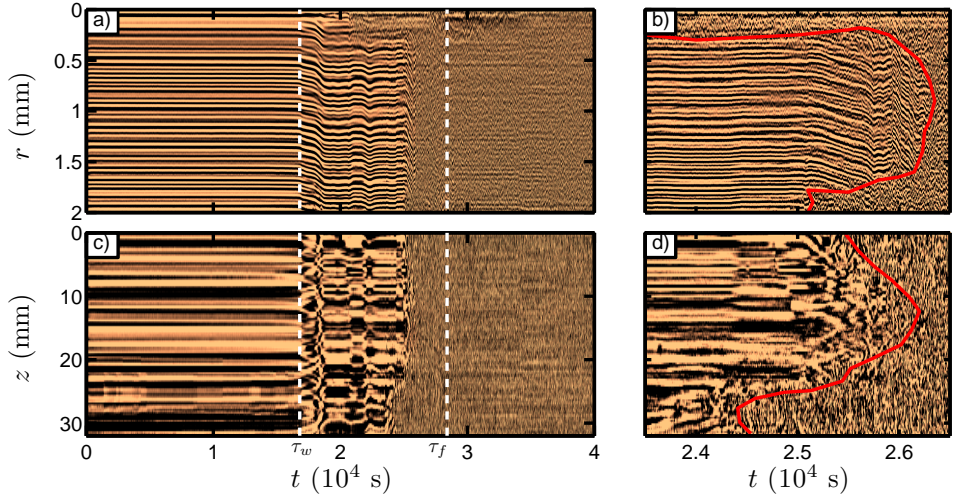}}
\caption{Ultrasonic imaging in a $6\%$~wt. carbon black gel under a constant oscillatory stress of amplitude $\sigma_0=11$~Pa at frequency $f_0=1$~Hz. The figure displays spatiotemporal diagrams of the speckle images $S(r,z,t)$ coded in linear color levels and recorded simultaneously to the rheological data shown in Fig.~\ref{fig2}.  See also Supplementary Movie~2 for an animation of the speckle images. (a-b)~$S(r,z_0,t)$ at depth $z_0=15$~mm from the top of the transducer array (corresponding roughly to the middle height of the Couette cell). (c-d)~$S(r_0,z,t)$ at horizontal position $r_0=1$~mm from the inner rotating cylinder (corresponding to the middle of the gap). (b) and (d) are enlargements of (a) and (c) within the fluidization regime $\tau_w<t<\tau_f$. The ultrasonic sampling frequency is equal to $f_{\rm US}=0.2$~Hz which is a submultiple of the oscillation frequency $f_0=1$~Hz. White dashed lines in (a) and (b) indicate $\tau_w$ and $\tau_f$. Red lines in (c) and (d) indicate the boundary between solidlike and fluidlike regions.}
\label{fig8}
\end{figure}

Figure~\ref{fig8} shows two spatiotemporal diagrams of the ultrasonic speckle images $S(r,z,t)$ recorded during the same experiment as that analyzed in Figs.~\ref{fig2}--\ref{fig7} above with a sampling frequency $f_{\rm US}=0.2$~Hz, i.e., ultrasonic images are taken every 5 cycles at the same intra-cycle time (see also Supplementary Movie~2). The color levels correspond to the image intensity $S$ at a fixed vertical position $z_0$ in the middle of the Couette cell [Fig.~\ref{fig8}(a-b)] and at a fixed position $r_0$ in the middle of the gap [Fig.~\ref{fig8}(c-d)]. The distance $r$ is measured from the inner bob while $z$ is measured from the top of the transducer array. In both cases, the horizontal axis corresponds to the time $t$ elapsed since the application of oscillatory stress and white vertical dashed lines indicate the two times $\tau_w$ and $\tau_f$ inferred previously from the rheological data.

Three different regimes can be distinguished in Fig.~\ref{fig8}. (i)~For $t<\tau_w$, the horizontal lines in the spatiotemporal diagrams show that the sample comes back to the exact same position from one cycle to another. This indicates a reversible motion of the seeding particles consistent with an elastic response of the gel. (ii)~For $\tau_w<t\lesssim\tau_f$, solidlike response is lost and a complex, heterogeneous speckle pattern is observed. (iii)~Full decorrelation of the speckle pattern in both $r$ and $z$ directions is achieved for $t\gtrsim\tau_f$.

More precisely, for $t\gtrsim\tau_w$, the speckle traces in Fig.~\ref{fig8}(a) take up a constant slope throughout the gap. This is indicative of total slippage of the sample at the walls from one cycle to another. Therefore, apparent yielding at $t=\tau_w$ can be attributed to wall slip as a result of failure at the inner bob. This most probably explains the presence of even harmonics in $\gamma(t)$ for $t\simeq\tau_w$ as seen above in Fig.~\ref{fig4}. Then, for $t\gtrsim 2.1\,10^4$~s, the speckle pattern loses its temporal correlation for $r\lesssim 0.2$~mm. This means that acoustic scatterers close to the inner rotating wall move by a significant fraction of the ultrasonic wavelength (100~$\mu$m) within 5 oscillation cycles. We take this decorrelation as an evidence for irreversible motion of the scatterers and for {\it local} fluidlike behavior of the sample. As already discussed by \cite{Gibaud:2010}, irreversibility most likely corresponds to the sedimentation of the seeding particles within the fluidized material. For $r\gtrsim 0.2$~mm, the speckle signal keeps a clear temporal correlation up to $t\simeq 2.5\,10^4$~s [see Fig.~\ref{fig8}(b)]. This means that for about one hour, the sample remains locally solidlike in most of the gap and that this solid coexists with a small fluidized region close to the inner moving wall. As seen in Fig.~\ref{fig4}, even harmonics in the strain response become negligible when total wall slip gives way to solid--fluid coexistence at $t\simeq 2.1\,10^4$~s. For $t\gtrsim 2.5\,10^4$~s, fluidization also starts at the outer cup. The solidlike material in the bulk then gets progressively eroded from both ends by the surrounding fluidized suspension until full fluidization (i.e. full decorrelation of the speckle pattern) is achieved at $t\simeq 2.6\,10^4$~s at least for the specific value of $z_0$ considered in the spatiotemporal diagram of Fig.~\ref{fig8}(a,b).

\begin{figure} [t]
\centerline{\includegraphics[width=1\columnwidth]{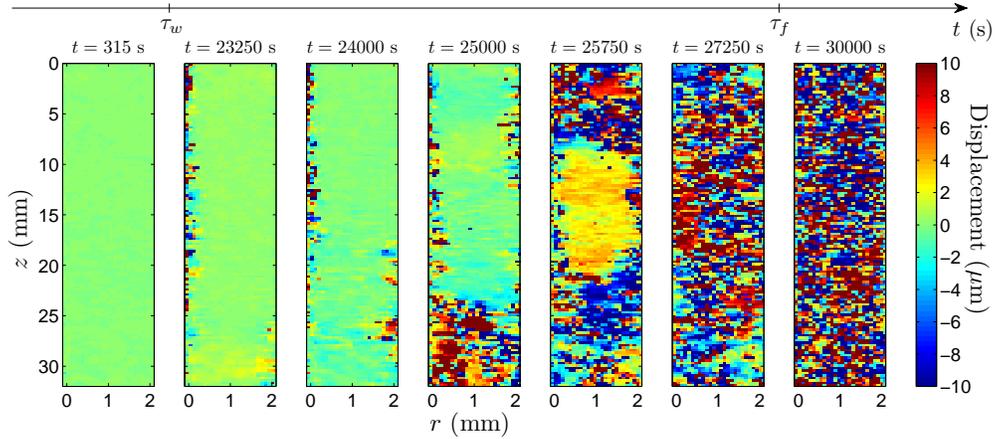}}
\caption{Images of the displacement field $\Delta(r,z,t)$ between two successive ultrasonic pulses sent every $\delta t=1/f_{\rm US}=5$~s at various times during the yielding process. The displacements are coded in linear color levels. Same experiment as in Figs.~\ref{fig2}--\ref{fig8}. See also Supplementary Movie~2.}
\label{fig9}
\end{figure}

This sequence of events is similar to that reported by \cite{Gibaud:2010} based on one-dimensional ultrasonic velocimetry at middle height of a Couette cell for a larger oscillation amplitude of $\sigma_0=15$~Pa. However, here for $\sigma_0=11$~Pa, wall slip is observed to occur after a long delay $\tau_w$ while it is observed right from the application of oscillatory stress at the larger amplitude. Moreover, the present two-dimensional experiments provide an additional view over the vertical direction: Fig.~\ref{fig8}(c-d) shows that solid--fluid coexistence also occurs along the vorticity direction although it starts later (around $t\simeq 2.4\,10^4$~s) than in the radial direction. We recall that Fig.~\ref{fig8}(c-d) corresponds to $r_0=1$~mm where solidlike behavior persists for longer times so that it is not surprising that the speckle remains vertically homogeneous for longer times. We also note that fluidization starts from the bottom of the Couette cell then at the top before propagating towards the middle of the cell.

Full fluidization in both $r$ and $z$ directions is observed for $t\simeq 2.6\,10^4$~s after which the speckle remains uncorrelated i.e. the whole sample has become fluidlike. This fluidization time is about 10\% shorter than $\tau_f\simeq 2.8\,10^4$~s inferred above from rheological data. However, the ultrasonic images analyzed in Fig.~\ref{fig8} only correspond to specific values of $z_0$ and $r_0$ and to a given vertical slice of our CB gel. In view of the radial and vertical heterogeneities, it is most probable that the sample is also heterogeneous in the azimuthal direction so that $t\simeq 2.6\,10^4$~s naturally underestimates the actual fluidization time of the whole three-dimensional sample. Moreover, the red solid--fluid boundaries in Fig.~\ref{fig8}(b,d) are only indicative and such boundaries would most likely change if we used a finer ultrasonic sampling frequency and a truly quantitative criterion for speckle decorrelation. We conclude that the timescale $\tau_f$ inferred from rheology alone corresponds to full three-dimensional fluidization of the CB gel. 

In order to directly confirm the fluidization scenario deduced from speckle images, Fig.~\ref{fig9} displays a few displacement maps computed by cross-correlating successive speckle images at different times during yielding (see Supplementary Movie~2 for the full data set). Figure~\ref{fig9} confirms that no significant motion is observed as long as  $t<\tau_w$ while for $\tau_w<t<\tau_f$, strongly heterogeneous displacement maps are reported, first through the existence of a thin fluidized band close to the inner bob and later through the growth of fluidized pockets from both cylindrical walls and from the bottom then from the top of the Couette cell. Finally, for $t\gtrsim\tau_f$, the displacement map appears as spatially erratic over the whole sampled volume. This simply translates the fact that speckle images in the fluidized state are decorrelated and that irreversible motions from one oscillation cycle to another are too large to be reliably captured by our cross-correlation algorithm.

\section{Discussion and conclusions}
\label{s.discuss}

\subsection{Two timescales in the fatigue and fluidization of CB gels under LAOStress}

The most striking result of the present paper is the existence of two well-defined timescales $\tau_w$ and $\tau_f$ that characterize the time-dependent yielding of CB gels under LAOStress. The experiment investigated in details in Sections~\ref{s.cst} and \ref{s.usv} shows that these timescales can reach several $10^4$~s. This explains why a standard amplitude sweep cannot be used to estimate reliably the LAOStress response of CB gels. In particular, these colloidal gels yield even for stress amplitudes well below the apparent yield stresses inferred from amplitude sweeps. When the stress amplitude is kept constant, rheological measurements coupled to ultrasonic imaging show that (i)~for $t<\tau_w$, the gel behaves as an elastic solid subject to a slow fatigue process, (ii)~it yields locally at the walls for $t\simeq\tau_w$ and subsequently shows apparent wall slip for $t\gtrsim\tau_w$, (iii)~for $\tau_w<t<\tau_f$, the sample is constituted of macroscopic fluidized regions that coexist with solidlike domains which get progressively eroded, and (iv)~for $t>\tau_f$, the sample is eventually fully fluidized and rheological data are dominated by inertia.

Interestingly, recent creep experiments by \cite{Grenard:2013pp} on the same CB gels in a roughened Couette geometry have also revealed the existence of two different timescales, noted $\tau_c$ and $\tau_f$, associated with a similar scenario. Under a steady shear stress $\sigma$, the material was shown to first slowly creep until it fails at the inner bob at $t=\tau_c$ and then to get progressively and heterogeneously fluidized until $\tau_f$. Such a striking similarity between creep and LAOStress suggests that the same delayed yielding mechanism is at play under both constant stress and oscillatory stress. We also expect this scenario to be relevant to other soft materials showing delayed yielding under creep such as thermo-reversible gels [\cite{Gopalakrishnan:2007,Brenner:2013}], silica gels or depletion gels [\cite{Sprakel:2011}]. The general picture involves failure at the walls and subsequent erosion of the solid material into macroscopic pieces surrounded by larger and larger fluidized regions. This picture is reminiscent of that found in Laponite suspensions under steady shear yet under smooth boundary conditions only \cite{Gibaud:2009}. Clearly, the influence of surface roughness on the yielding processes under LAOS should also be addressed in future work.

\subsection{Local vs global characterization of time-dependent yielding under LAOStress}

In the course of this paper, we have highlighted the interest of both Fourier transform rheology and LAOStress measures to characterize a time-dependent material under an oscillatory stress. Our results also reveal important limitations of such approaches based on global rheological data alone. Indeed, ultrasonic imaging has allowed us to evidence strong spatial heterogeneities for $\tau_w<t<\tau_f$ in both gradient and vorticity directions. Such heterogeneity questions the interpretation (if not the validity) of the spatial averages provided by the rheometer. Obviously, in the presence of slippage at the walls, the global strain $\gamma$ only gives an indication on the oscillation amplitude and is not representative of the actual strain within the material. 

\begin{figure} [h]
\centerline{\includegraphics[width=1\columnwidth]{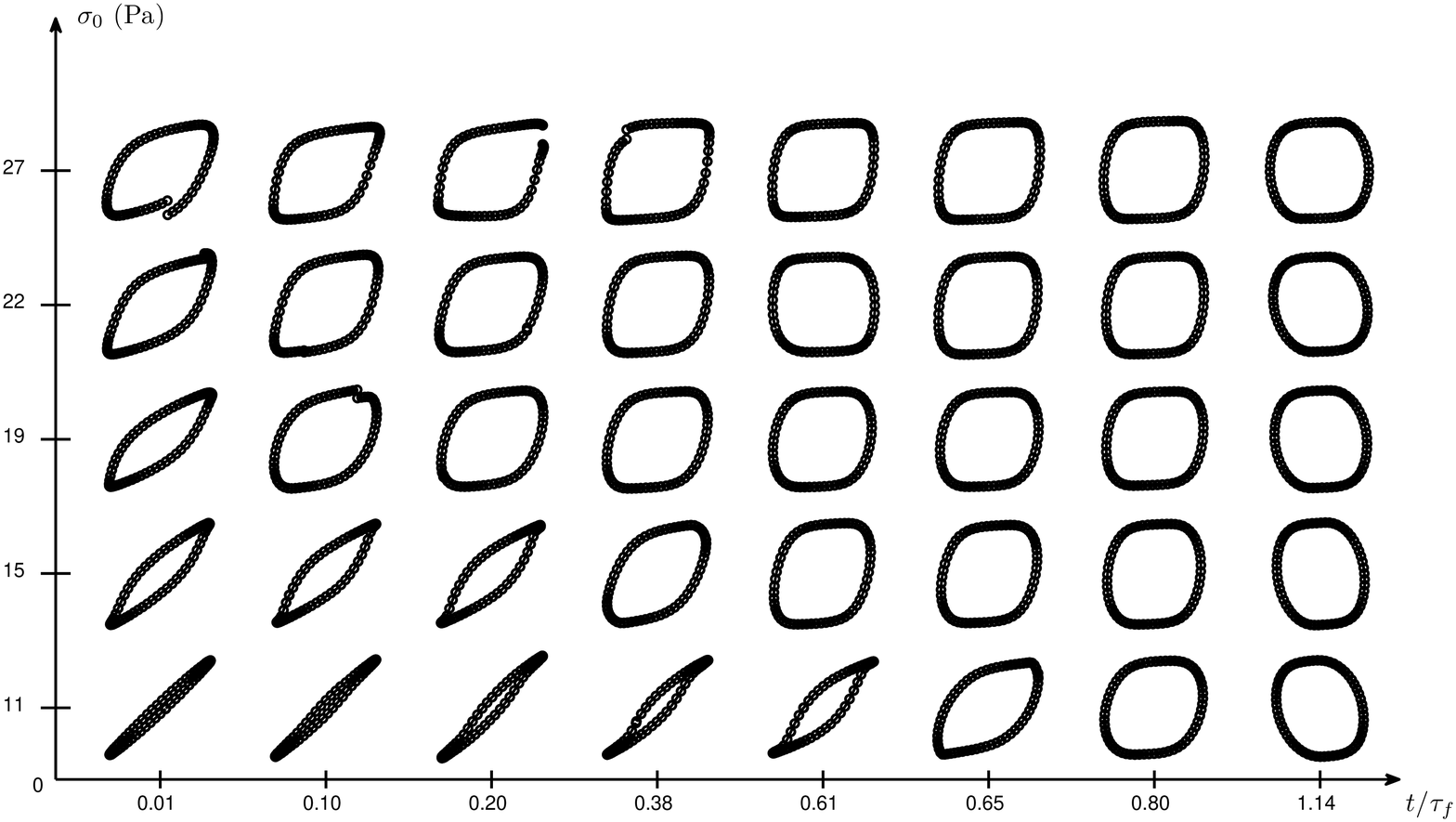}}
\caption{Pipkin-like diagram of the time-dependent response of a CB gel to LAOStress: Lissajous plots ($\sigma$, $\gamma$) at different times $t/\tau_f$ in the yielding process and for different amplitudes $\sigma_0$ of the imposed oscillatory stress. The oscillation frequency is fixed to 1~Hz in all cases.}
\label{fig10}
\end{figure}

So far, we have only focused on a single experiment at $\sigma_0=11$~Pa. In view of the complexity of the fatigue and yielding scenario, a systematic investigation as a function of $\sigma_0$ is out of the scope of the present paper. Still, it is interesting to show some global rheological data for increasing values of $\sigma_0$ is the form of the Pipkin-like diagram drawn in Fig.~\ref{fig10}. Lissajous plots are gathered as a function of time $t$ (instead of frequency $\omega_0$ in the classical Pipkin representation) and of the stress amplitude $\sigma_0$. The data investigated so far are shown in the bottom line. The temporal evolution of the Lissajous plots is seen to change sharply upon increasing $\sigma_0$. This indicates that the ratio $\tau_w/\tau_f$ is not a constant and strongly depends on $\sigma_0$. In particular, the elastic behavior is no longer observed for stresses larger than 11~Pa and for $t/\tau_f=0.01$. This is due to the fact that at larger stresses wall slip occurs as soon as the stress is applied in agreement with \cite{Gibaud:2010}. Moreover, the fact that Lissajous plots are not closed loops for the highest stress amplitudes and for the shorter times clearly points to a time dependence of the material on the timescale of one cycle (here 1~s). A detailed study of the local two-dimensional yielding scenario at stresses larger than 11~Pa and of the dependence of the two timescales $\tau_w$ and $\tau_f$ on $\sigma_0$ is part of ongoing work.

\subsection{Open questions and perspectives}

To the best of our knowledge, the present experiments provide the first direct evidence for successive failure at the walls and heterogeneous bulk fluidization of a colloidal gel under LAOStress. Our macroscopic interpretation of the two timescales $\tau_w$ and $\tau_f$ based on ultrasonic imaging obviously differs from the microscopic picture based on bond stretching and cage breaking proposed for the two characteristic yield strains found in attractive glasses under both LAOStrain and constant strain [\cite{Pham:2006,Laurati:2011}]. Whether or not it is possible to reconcile these different pictures stands out as an open issue. In particular, time-resolved LAOS experiments (instead of amplitude sweeps) close to yielding coupled to local measurements at scales larger than the individual colloids would be very interesting to confirm the steady-state interpretation for attractive glasses. Similar local measurements are also in line to show possible spatial coexistence of the soft and solid states recently evidenced in the oscillatory yielding of glassy star polymers [\cite{Rogers:2011b}].

More generally, a theoretical framework for LAOS is currently lacking for time-dependent materials. Such a framework is however required to understand at least the global data shown in Fig.~\ref{fig10}. While including the effects of boundaries and slippage at the walls still seems out of reach, models that account for some degree of thixotropy and heterogeneity under LAOS are underway. For instance, the recent isotropic-kinematic hardening model devised by \cite{Dimitriou:2013pp} should be very fruitful in interpreting experimental data on time-dependent materials, although its application to delayed yielding in colloidal gels remains questionable. 

Finally, intra-cycle information based on ultrasonic imaging is also technically available since the temporal resolution of the setup described in Sect.~\ref{s.usvsetup} can reach 10,000 frames per second. Therefore, future work will not only focus on investigating more systematically the influence of the stress amplitude but also on a spatially and temporally resolved study of the evolution of the strain field within one cycle. 

\begin{acknowledgments}
The authors wish to thank J{\"o}rg L{\"a}uger at Anton Paar as well as Aloyse Franck and Raoul Smith at TA Instruments for technical advice on data acquisition and analysis. We also thank Vincent Grenard for preliminary experiments and Thibaut Divoux, Gareth McKinley, and Stefan Lindstr{\"o}m for fruitful discussions. This work was funded by the Institut Universitaire de France and by the European Research Council under the European Union's Seventh Framework Programme (FP7/2007-2013) / ERC grant agreement No.~258803. TG acknowledges funding from the french Agence Nationale de la Recherche (grant No.~ANR-11-PDOC-027). 
\end{acknowledgments}


\end{document}